# Detecting short period variable stars with Gaia

Mihaly Varadi[1], Laurent Eyer[1], Stefan Jordan[2], Nami Mowlavi[1], Detlev Koester[3]

*1 – Geneva Observatory, University of Geneva, ch. des Maillettes 51, CH-1290 Sauverny, Switzerland*
*2 – ARI/ZAH, Univ. of Heidelberg, Mönchhofstr. 12-14, D-69120 Heidelberg, Germany*
*3 – Institut für Theoretische Physik und Astrophysik, University of Kiel, Leibnizstraße 15, D-24098 Kiel, Germany*

**Abstract.** We analyzed the frequency domain of time series of simulated ZZ Ceti light-curves to investigate the detectability and period recovery performance of short period variables (periods < 2 hours) for the Gaia mission. In our analysis, first we used a non-linear ZZ Ceti light-curves simulator code to simulate the variability of ZZ Ceti stars (we assumed stationary power spectra over five years). Second we used the Gaia nominal scanning law and the expected photometric precision of Gaia to simulate ZZ Ceti time series with Gaia's time sampling and photometric errors. Then we performed a Fourier analysis of these simulated time series. We found that a correct period can be recovered in ~65% of the cases if we consider Gaia per CCD time series of a G ~ 18 magnitude multiperiodic ZZ Ceti star with 5%-10% light-curve variation. In the pre-whitened power spectrum a second correct period was also recovered in ~26% of the cases.

## INTRODUCTION

The ESA Gaia satellite will observe about one billion stars with unprecedented astrometric and photometric precision. Over its 5 year mission, it will systematically scan all the sky and observe sources from 40 to 250 times, down to magnitude G ~ 20 mag. See info-sheet at http://www.rssd.esa.int/Gaia for a more detailed description of the satellite.

**TABLE 1.** Types and properties of short period variables.

| Type | Periods [minutes] | Amplitudes [mag] |
|---|---|---|
| β Cep stars | 96 - 480 | < 0.1 |
| δ Scuti stars | 28 - 480 | 0.003 - 0.9 |
| roAp stars | 6 - 21 | < 0.01 |
| EC14026 stars | 1.3 - 8.3 | < 0.03 |
| Betsy stars (PG1716) | 33 - 150 | < ~0.01 |
| ZZ Ceti stars (DAV) | 0.5 - 25 | 0.001 - 0.3 |
| V777 Her stars (DBV) | 2 - 16 | 0.001 - 0.2 |
| GW Vir stars (DOV + PNNVs) | 5 - 85 | 0.001 - 0.2 |
| Eclipsing white dwarfs | > 6 | < 0.7 |

The Gaia time sampling and the CCD data acquisition scheme allow in principle to probe stellar variability on time scales even as short as several tens of seconds, thereby giving potential access to the study of variable stars in a large and homogenous sample of stars. A first study was done by Eyer & Mignard (2005) on the correct detection rate of monoperiodic signals for a wide range of periods. They concluded that periods of regular variable star can be recovered even from signals with low S/N ratio and that the period recovery depends mainly on the ecliptic latitude. Later Mary et al. (2006) presented a work on the detectability of low amplitude short period multiperiodic signals, which correspond to pulsation modes of roAp stars. Our goal is to extend these studies using more realistic light-curve models for several types of variables stars (see Table 1), which are showing variability on timescales less than 2 hours. We call these short period variables. The variability amplitudes of short period variables are mostly at millimagnitude level, so that a good photometric precision is a key point to the detection of such kind of stars. Gaia's design seems to meet this requirement (see Fig. 1 for the expected photometric precision of Gaia).

In this article we are analyzing simulated Gaia time series of ZZ Ceti variables. ZZ Ceti stars are multiperiodic white dwarf pulsators (DAVs), which are oscillating in low order non-radial gravity modes (see e.g. Montgomery in these proceedings). From the point of view of our analysis, it is worth to note, that most of the ZZ Ceti stars show amplitude, period and mode changes on timescales from weeks to years, which causes that their power spectra are not stationary over the mission life time of Gaia. The study of such light-curves is not part of this work, but we plan to investigate the detectability of variables which do not have stationary power spectra.

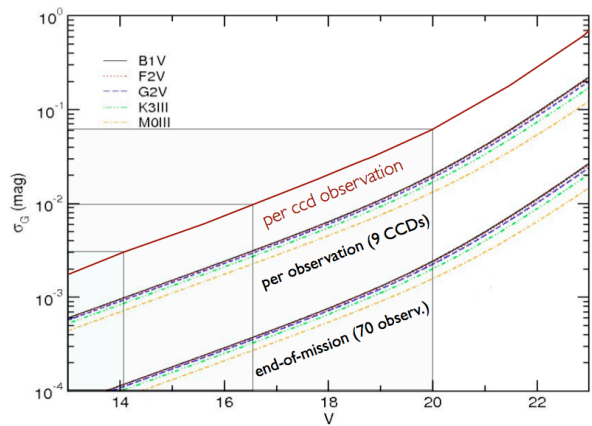

**FIGURE 1**. Expected photometric precisions of Gaia: per-CCD, per transit, and over the whole mission. In the last two cases the dependence on spectral types is also shown. Only the photon noise is taken into account, but calibration uncertainties will dominate the bright end. Figure by C.Jordi (private communication) modified by M.Varadi.

## LIGHT-CURVE SIMULATIONS

To perform tests on the detectability of ZZ Ceti stars from Gaia like photometry, first we simulate continuous ZZ Ceti light-curves, second we generate Gaia time series from these by taking into account the time sampling and photometric precision of the Gaia satellite.

## ZZ Ceti light-curves simulation

We use a nonlinear ZZ Ceti (DAV) light-curve simulator based on a model originated from D.Koester and provided by S.Jordan. The model and the simulator code are described in the diploma thesis of Schlundt (2006). The Fortran code was rewritten in C++ by M.Varadi and was made suitable for sparse and long Gaia light-curve simulations.

The model takes periods, amplitudes and phases as input to define a pulsation pattern at the base of the convective zone and solves the problem of flux propagation through the convective zone to derive the relative flux variations at the photosphere. Examples are shown in Fig. 2. Similar models were successfully fitted to mono and multiperiodic white dwarf pulsators e.g. by Montgomery (2005, 2007).

To simulate the continuous light-curves presented in Fig. 2 we took the periods, amplitudes and phases of the multiperiodic ZZ Ceti star GD29-38 and feed this data as input to our ZZ Ceti light-curve modeling code described above.

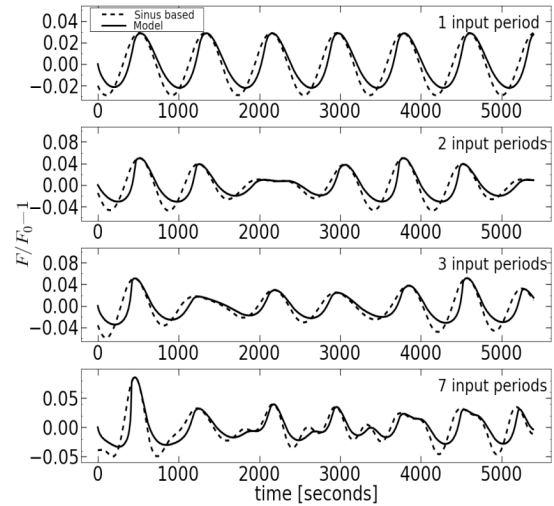

**FIGURE 2**. Simulated ZZ Ceti light-curves. The dashed curves are the input signals at the base of the convective zone. Each of them is a sum of sinusoids with different periods. The number of periods is indicated in each panel. The continuous curves are the corresponding simulated ZZ Ceti light-curves, as it would be observed in the photosphere.

## Gaia time series simulation

In order to simulate photometric Gaia time series of ZZ Ceti stars, first we sample the continuously simulated ZZ Ceti light-curves according to the nominal scanning law of Gaia, second we add Gaussian noise, which follows the photometric precision of Gaia shown in Fig. 1. There are three software packages available to generate Gaia time sampling for a given position on the sky. For the simulations of this work we used the AGISLab package developed by ELSA[1] fellow B.Holl (see Holl et al. 2009).

### Simulation Parameter space

We chose 100 different positions on the sky. All of them correspond to 80 field-of-view transits and 720 per-CCD measurements. This is the average number of measurements of a source at end of mission. We used per-CCD photometry because previous tests showed higher period recoverability percentages with respect to field-of-view transit photometry (see Lecoeur et al. 2009).

---

[1] ELSA – is a Marie Curie Research Training Network built around Gaia. http://www.astro.lu.se/ELSA/index.html

The SDSS DAVs sample magnitude distribution has a mean around 18 mag. Therefore we chose a G ~ 18 magnitude star for which Gaia will have ~20 mmag per-CCD photometric errors according to Fig 1. We generated 5 year long time series with stationary power spectra. We used a signal with 7 frequency components to describe the pulsation at the base of the convective zone the three main frequencies are given in Table 2. With this choice of amplitudes we have a signal-to-noise ratio close to the unity.

**TABLE 2. The three main frequencies (out of 7) used as input to build the signal in the base of the convective zone.**

| Periods [seconds] | Amplitudes [mi] | Phases [degree] |
|---|---|---|
| 816 | 0.0288 | 224 |
| 653 | 0.0211 | 165 |
| 614 | 0.0220 | 253 |

## ANALYSIS

Testing the detections of short period variability from Gaia photometry can be done in several ways. We chose the standard Fourier Transform (see Deeming 1975) with three pre-whitening steps in order to investigate the achievable multi-period recovery rates for the simulated non-linear ZZ Ceti light-curves. In each step of the pre-whitening cycle, we compare the frequency of maximum peak found in the power spectrum with the input periods used for simulating the signal (see previous section).

We found that one of the three main frequencies given in Table 2 can be recovered in ~65% of the cases for a G ~18 magnitude star. In the second step of the pre-whitening cycle we found that in ~26% of the cases a correct secondary period can also be recovered. For comparison we performed the same analysis with noiseless data. We found ~72% period recovery percentages for the first step and ~41% for the second step.

## CONCLUSIONS

We investigated the period recovery performances of simulated Gaia time series of ZZ Ceti stars. With the assumption that ZZ Ceti light-curves have stationary power spectra over five years, we found that the one of main periods can be recovered from simulated photometric Gaia data, in ~65% of the cases.

Usually ZZ Ceti stars with high amplitudes do not have stable light-curves. This is questioning our results. However, there are other types of multiperiodic short period variables in this period range where our test method of period recovery could be applicable with more confidence on the results. For example sub-dwarf variables (see Charpinet 2009 in these proceedings) show more stable light-curves over long period of time.

In the future we would like to test period recovery performances of non-stationary ZZ Ceti light-curves as well as other short period variables. Lastly it is also among our plans to find fast methods for detecting short period variables in the entire Gaia database because searching for very short periods within long and sparse Gaia time series for one billion objects is very CPU and time demanding even in the era of supercomputers.

## ACKNOWLEDGMENTS

We thank the Marie Curie Research Training Network ELSA, which is supported by the European Community's Sixth Framework Programme under the contract MRTN-CT-2006-033481. We thank Professor Daniel Pfenniger for the computing resources for the simulations. We also thank Ádám Sódor and Berry Holl for the help they provided with the software packages that we are using to carry out the simulations presented here.